\documentclass[preprint,12pt]{aastex}
\begin{document}
\title{The White Dwarfs within 25 Parsecs of the Sun: Kinematics and Spectroscopic Subtypes}                

\author{Edward M. Sion}\affil{Dept. of Astronomy \& Astrophysics,Villanova University,Villanova, PA 19085, e-mail: edward.sion@villanova.edu}

\author{J. B. Holberg}\affil{Lunar and Planetary Laboratory,University of Arizona,Tucson, AZ 75201,email: holberg@vega.lpl.arizona.edu}

\author{Terry D. Oswalt}\affil{Dept. of Physics and Space Sciences,Florida Institute of Technology,Melbourne, FL 19085,email: toswalt@fit.edu}

\author{George P. McCook}\affil{Dept. of Astronomy and Astrophysics,Villanova University,Villanova, PA 19085,email: george.mccook@villanova.edu}

\author{Richard Wasatonic}\affil{Dept. of Astronomy and Astrophysics,Villanova University,Villanova, PA 19085,email: richard.wasatonic@villanova.edu}

\author{Janine Myszka}\affil{Dept. of Astronomy and Astrophysics,Villanova University,Villanova, PA 19085,email: janine.myszka@villanova.edu}

\begin{abstract}

We present the fractional distribution of spectroscopic subtypes, range and distribution of surface temperatures, and kinematical properties of the white dwarfs within 25pc of the sun. There is no convincing evidence of halo white dwarfs in the total 25 pc sample of 224 white dwarfs. There is also little to suggest the presence of genuine thick disk subcomponent members within 25 parsecs. It appears that the entire 25 pc sample likely belong to the thin disk. We also find no significant kinematic differences with respect to spectroscopic subtypes. The total DA to non-DA ratio of the 25 pc sample is 1.8, a manifestation of deepening envelope convection which transforms DA stars with sufficiently thin H surface layers into non-DAs. We compare this ratio with the results of other studies. We find that at least 11\% of the white dwarfs within 25 parsecs of the sun (the DAZ and DZ stars) have photospheric metals that likely originate from accretion of circumstellar material (debris disks) around them. If this interpretation is correct, then it suggests the possibility that a similar percentage have planets, asteroid-like bodies or debris disks orbiting them. Our volume-limited sample reveals a pileup of DC white dwarfs at the well-known cutoff in DQ white dwarfs at $T_{eff} \sim $6000K. Mindful of small number statistics, we speculate on its possible evolutionary significance.  We find that the incidence of magnetic white dwarfs in the 25 pc sample is at least 8\%, in our volume-limited sample, dominated by cool white dwarfs. We derive approximate formation rates of DB and DQ degenerates and present a preliminary test of the evolutionary scenario that all cooling DB stars become DQ white dwarfs via helium convective dredge-up with the diffusion tail of carbon extending upward from their cores. 

\end{abstract}

Subject Headings: techniques: photometric - stars: techniques:spectroscopic - stars: white dwarfs - stars: statistics - stars:kinematics

\section{Introduction}

Extending the census of local white dwarfs to increasingly large volumes of space around the sun offers a plethora of crucial astrophysical insights
including (a) identification of the lowest luminosity, hence oldest, white dwarfs in the immediate neighborhood of the sun; (b) the distribution of white dwarf spectroscopic subgroups and their mix of progenitor stellar populations; (c) a direct measurement of the local space density and mass density of white dwarfs and; (d) a window into the history of star formation and stellar evolution in the Galactic plane
as well as constraining the age of
the Galactic disk (Liebert, Dahn, Monet 1988, Oswalt et al. 1996) by enabling the determination of the cool white dwarf luminosity function. The sample of WDs out to 20 pc was presented by Holberg et al. 2008) and Sion et al. (2009) who utilized available
temperatures, gravities, and spectral types from the literature to attempt a full characterization of the local sample of white dwarfs. With the publication of new atmospheric parameters, more accurate distances for a larger sample of local white dwarfs out to 25 pc, the statistical and kinematical properties of the volume-limited sample of local white dwarfs merit re-examination.

The local 25 pc white dwarf sample is a well characterized, volume-limited population with a high degree of completeness.
It contains a sample of 224 white dwarfs including double degenerate systems. While it has an overall completeness of 65\%, the 20 pc and 13 pc samples have completeness's of 85\% and 100\%, respectively. As such it lends itself to 
a number of analyses that can be used to anchor larger, less complete spectroscopic and photometric surveys which are not volume-limited.  The results of the current 25 pc sample will be discussed in Holberg et al.(in preparation) which will include an evaluation of the spatial distribution, the space density and completness of the sample as well as the mass distribution and the local luminosity function and binary fraction.
 
In the present paper, we discuss the kinematics, the distribution of white dwarf temperature and spectroscopic subtypes in the 25 pc sample around the sun and propose tests of scenarios of cool white dwarf spectral evolution. We briefly describe the enhanced 25 pc sample in Section 2.  The kinematic properties of this sample are discussed in Section 3.  The DQ and DC components of the 25 pc sample and their temperature distributions are discussed in Section 4.  The magnetic and cool DZ stars are discussed in Sections 5 and 6 respectively.  In Section 7 we describe preliminary tests of cool white dwarf spectral evolution using the volume-limited 25 pc sample.  Section 8 contains our conclusions.

\section{Census of White Dwarfs Out to 25 pc}

Table 1 presents the sample of white dwarfs within 25 pc of the sun. The basic observational data from which we examined the distribution of white dwarf spectral types and the data used for space motions, are given in Table 1, which contains by column: (1) the WD-number, (2) the coordinates (RA and DEC are in decimal degrees), (3) $T_{eff}$, (4) DIST (distance in parsecs), (5) PM (proper motion in arcsec/yr), (6) PA (position angle in degrees), (7) the method of distance determination, denoted by $p$ for trigonometric parallax, $s$ for spectrophotometric distances and (8) a reference abbreviation to the bibliography.  Many of the distances are obtained from trigonometric parallaxes, the remainder from spectrophotometric distances. 

A substantial fraction of the data in Table 1 was taken from the detailed survey of the local white dwarfs out to 20 parsecs by Holberg et al.(2008) and Sion et al. (2009). To this sample of 129 white dwarfs, we have added 95 white dwarfs out to a volume of radius 25 pc, giving a total sample of 224 white dwarfs including double degenerates known out to 25 pc. As in Holberg et al. (2008) and Sion et al. (2009), the photometric distances were computed based upon spectroscopic and photometric measurements following the methods described in Holberg, Bergeron \& Gianninas (2008). Proper Motions are taken from the McCook and Sion Catalog, or where available, were determined from NOMAD. Radial velocities were available from the literature for approximately 50\% of our sample, and correspond either to direct measurements of the white dwarf or the system velocities or radial velocities of the main sequence companions (McCook \& Sion 1999 and references therein; Maxted et al.2000, Silvestri et al. 2001, 2002, Pauli et al. 2003, 2006). For radial velocities derived from individual white dwarfs, we applied corrections for the gravitational redshift based on the individual masses and radii of each star.  These mass and radius determinations were interpolated from within the synthetic photometric tables described in Holberg \& Bergeron (2006) and were based on temperatures and gravities given in LS08.  Using the spectral types and temperatures given in Holberg et al.(2008,hereafter, LS08) and Holberg (2008), we assembled the local population of white dwarfs lying within 25 pc of the sun, using a number of compilations including Gliese \& Jahreise (1995), LS08, Sion et al. (2009), Sayres et al. (2012), Bergeron et al. (2011),	 Giamichele et al. (2012), Dufour et al. (2007), Limoges et al. (2013), Subasavage et al. (2007, 2008), Gianninas et al. (2011), Kilic et al. (2010), Koester et al.(2011), and Tremblay et al. (2011).	 					   
				   
We summarize the percentage breakdown of spectral subtypes among the 25 pc sample of local white dwarfs in Table 2.  As expected, the DA stars dominate the sample. If we take the total DA sample to include DAZ stars and magnetic DA stars, then there are 150 DA stars within 25 pc.

In Fig.1 we display a histogram, with the number versus $T_{eff}$ distribution function of DA stars in the lower panel and non-DA stars in the upper panel. The skewing of their distribution toward lower temperatures not only reflects the predominance of cool DAs and cool non-DAs in a sample so close to the sun but the variation of this ratio, DA/non-DA, as a function of $T_{eff}$, holds key physical significance to our understanding of white dwarf spectral evolution. 

The overall DA to non-DA ratio of the 25 pc sample is 1.83. 
This is a slightly larger ratio than found in the 20 pc sample and is likely a lower limit since a few of the cool non-DAs (especially DC stars) may actually prove to be H-dominated. 
Sion (1984) uncovered the first empirical evidence of this  transformation of white dwarf surface compositions and attributed the lowering of the DA/non-DA ratio with decreasing $T_{eff}$ to the mixing and dilution of hydrogen into the deepening helium convection zone (Sion 1984; see also Greenstein 1986). Sion (1984) estimated that this convective mixing away of surface  hydrogen in cooling DA stars occurs at $T_{eff}$ $\sim$ 10-12,000K. Subsequent more detailed work with evolutionary models (Tremblay and Bergeron 2008; Bergeron, Ruiz \& Leggett 2001), using a larger, more homogenous sample of white dwarfs, presented a more complex picture of spectral evolution attemperatures below 15000K. They showed that 
as a DA white dwarf cools, the bottom of the hydrogen convection
zone eventually reaches the helium convection zone but only if the hydrogen layer is thin enough, with the actual temperature at which the conversion occurs, depending on the thickness of the hydrogen layer. In other words, DA White dwarfs with thicker hydrogen layers will mix at lower $T_{eff}$. They showed that the H layer mass that would mix at a given temperature, depends upon the assumed convective efficiency, with $\sim$85\% of all DA stars between 10,000K and 15,000K having hydrogen layer masses large enough for them to remain recognized as DA stars down to 8000K. It is clear from the results of Tremblay and Bergeron (2008) that the details of white dwarf spectral evolution remain murky below 12000K. This can only be clarified with further theoretical modeling predictions confronted by ever larger samples of cool white dwarfs with high quality spectra.

\section{Kinematics of the Local White Dwarfs}

The vector components of the space motions are U, V, and W, where U is positive in the direction of  the galactic anti-center, V is positive in the direction of the galactic rotation and W is positive in the direction of the north galactic pole (Wooley et al. 1970). The U, V, and W components and total motions were computed along with the average velocities and velocity dispersions for the different spectroscopic subtypes in the 25 pc sample in Table 1. In Table 3 below, we present the computed vector components of the space motion for each white dwarf in Table 1. We note that for some of the WDs in Table 1, there are two sets of space motions from slightly different input parameters, such as proper motion and position angles. For these objects, two different sets of U,V,W,T values were computed and used to compute the averages and dispersions tabulated in Table 4. These mean velocities and velocity dispersions reveal no clear evidence that different spectral subtypes have significantly different space motions.

If we take $T > 150$ km/s as the lower cutoff for halo space motions, then there are seven stars or 3\% of the total sample that have possible halo population II stars in extreme galactic orbits that could happen to be interlopers in the solar neighborhood. These objects are WD0747+482.1 (DC10.4; 156 km/s),WD0747+482.2 (DC12.0; 156 km/s), WD0959+149 (DC7; 144 km/s), WD1310-472 (DC11.9; 166 km/s), WD1334+039 (DA11.0; 151 km/s), WD1339-340 (DA9.5; 215 km/s) and WD1756+827 (DA6.9; 156 km/s). 

To investigate the question of population membership further,
we point out that the assignment of reliable population membership for local white dwarfs must involve not only the vector components of the space motion but also the cooling ages derived from their surface temperatures and total stellar age.
In contrast, the population membership of main sequence stars utilizes kinematical charateristics as well as chemical abundance data, e.g. metallicity. Despite this difference in the way population membership is assigned for the two types of stars, it is useful to compare the velocity distribution of the white dwarf sample in the U versus V velocity plane with the velocity distribution (velocity ellipsoids) characterizing a well-studied sample of main sequence stars (Chiba \& Beers 2000, Vennes \& Kawka 2006, Sion et al. 2009).

However, a halo or thick disk member star cannot be 
identified on the basis of kinematic data alone. The candidate star must also have a total stellar age that is of order 12 billion years or older. The temperature of the six stars given in Table 1 indicate cooling ages well below 12 billion years. Thus, their space motions together with their total stellar ages, lead to the conclusion that there is no clear evidence of halo white dwarfs among the white dwarfs within 25 pc of the sun.

In Fig. 2, we display the U versus V space velocity diagram for the 25 pc sample of white
dwarfs, with the assumption of zero radial velocity, relative to velocity ellipses for main
sequence stars (Chiba \& Beers 2000; see also Kawka \& Vennes 2006, Sion et al. 2009). DA stars are denoted by closed circles, while non-DA stars are denoted by closed triangles. In the diagram, the $2 \sigma $ velocity ellipse contour (denoted by the solid line) of the thin disk component is displayed, as well as the $2 \sigma$ ellipse of the thick disk component (short dashed line) and the $1 \sigma$ contour of the halo component (long dashed line). The vast majority of white dwarfs lie within the thin disk as expected for the local sample. However, we see that nearly equal small numbers of DAs and non-DAs lie outside the thin disk ellipsoid but within the thick disk ellipse, while three stars, two DAs and one non-DA lie clearly within the halo velocity ellipse. Finally, one non-DA and two DA stars have anomalous velocities placing them at very large positive V velocity components. 

Given the concentric, nested nature of the thin and thick disk velocity ellipses in the U-V plane, it is difficult to clearly disentangle the two populations.  Age is a strong discriminate and careful consideration of this together with the use of individual galactic orbits (see Pauli et al. 2006 and others) may help to better separate these populations.

\section{DQ Stars within 25pc}

In Fig.3, we display the distribution of $T_{eff}$ for the 23 DQ white dwarfs (upper panel) and 28 DC white dwarfs (lower panel) within 25 pc. Although any comparison with the non-DA subgroups suffers from small number statistics, the number of DQ stars appears to peak at lower temperatures as expected since the $C_2$ molecular absorption bands strengthen with decreasing $T_{eff}$.  Although the number of stars in both distributions is small, it is noteworthy that the DC stars extend to even lower temperatures than the DQ stars. The previously known  precipitous cutoff in the number of DQ stars at surface temperatures cooler than $\sim$ 6500 is seen in Fig.3. This real cutoff was first noted by Bergeron, Ruiz and Leggett (1997) and Bergeron, Leggett and Ruiz (2001) and subsequently found in the much larger SDSS sample of DQ stars by Dufour et al. (2005) and by Koester \& Knist (2006). 

Returning to the temperature distribution of 28 DC stars and 23 DQ stars, while mindful of small number statistics, it is nonetheless interesting that we see a large increase in the number of DC stars beginning at or near the temperature at which the DQ cutoff is seen. While some of these DC stars may show line features at high enough spectral resolution and sensitivity while other DCs may turn out to be hydrogen-dominated, we suggest the possibility that some of the 28 DCs within 25 pc could possibly be heretofore unidentified DQ stars with extreme pressure shifts that render their spectra unrecognizable. Alternatively, there may be some physical mechanism that removes carbon from the smallest optical depths in the atmosphere.

The peculiar DQ stars, the so-called 
$C_2H $ stars, appear at the low temperature end of the  
DQ distribution of $T_{eff}$. Hall \& Maxwell (2008) pointed out that the peculiar absorption features first identified as $C_2H$
in the spectra of cool ($T_{eff} < 6000$K) peculiar DQ stars were misidentified as hydrocarbon molecules and instead are the  Swan bands of of $C_2$ that are extremely pressure-shifted. Kowalski (2010) showed that the distortion of Swan bands originates in the pressure-induced increase in the electronic transition energy between states involved in the transition. Unfortunately, the predicted Swan band shifts are too large compared to the observed ones. Thus, the need for further work in this area is indicated. 

\section{Magnetic White Dwarfs within 25 pc}

The true incidence of magnetic white dwarfs in a volume-limited survey remains an open question. Out to 25 pc, we count a total of 19 magnetic degenerates among which there are fourteen DA magnetics and five non-DA magnetics.
Liebert et al. (2003) presented evidence that the true incidence of detected magnetism among field white dwarfs at the level of $\sim$ 2MG or greater, is at least 10\%, and may be higher. The incidence of detected magnetism in our volume-limited sample is 8\% but would be as high or higher than the Liebert et al. (2003)
percentage if surveys to field strengths below
2 MG are carried out for our sample. 

The relatively small number of magnetics within 25pc
prevents a determination of whether or not the field strength of a magnetic white dwarf varies as a function of time (stellar age). For the same reason, we cannot say if the fraction of magnetic white dwarfs is higher among cool star samples (e.g. samples of field white dwarfs near the sun) than among surveys of field white dwarfs that extend out to greater volumes of space and hence sample hotter WDs. It remains an open question whether the number of magnetic white dwarfs increases with decreasing $T_{eff}$, luminosity or cooling age. The answers to these questions must await volume-limited surveys sufficiently large in radius to encompass substantial numbers of hot white dwarfs across a broader range of $T_{eff}$.

\section{Cool White Dwarfs with Metals}   

We count 26 cool white dwarfs in the sample which exhibit absorption features due to accreted metals. This number excludes the DQ stars because although some may accrete metals, the carbon in their atmospheres is primarily due either to convective dredge-up (Dufour et al. 2005) or is primordial (Dufour et al. 2007). There has been no definitive explanation for why the DZ stars and DQ stars appear to have mutually exclusive spectra; almost no DQ stars reveal absorption features due to metals (other than carbon) while the DZ stars rarely exhibit absorption features due to carbon.  
It is unlikely that accreted metals may be easier to hide in DQ atmospheres due to increased opacity provided by carbon. 
We note that Weidemann \& Koester (1989) analyzed a DZ star which contained carbon. More recently, Koester et al.(2011) reported a DQ star from the SDSS that reveals a decrease in flux at the blue end which could be explained with calcium and possibly iron. It is also possible that the apparent dichotomy between DQ and DZ stars could be due to selection and small number statistics. Moreover, since the DQ stars require very deep convection zones to dredge up carbon, it is also possible that any accreted metals are diluted to such an extent
that metal features are two weak to be detected.

Among these 26 white dwarfs within 25 pc are two DZAs, 13 DAZs and 11 DZ stars. For these objects, the most likely source of the accreted metals is from debris disks (Farihi et al. 2010 and references therein). 

On this basis, we speculate that all 26 non-DAs with photospheric metals have accreted from dust/debris disks of tidally disrupted asteroids, or in the case of DAZ stars, from volatile-rich matter such as comets. Thus, we speculate that at least 11\% of the white dwarfs within 25 pc have circumstellar debris disks with rocky, metallic debris and have very likely descended from main sequence progenitors that had planetary systems. By comparison with frequencies of occurrence of exoplanets among sun-like stars, Howard (2013) and Mayor et al.(2011) find that 15\% of Sun-like stars host one or more planets with mass Msini = 3-30 earth masses orbiting within 0.25 AU, and another 14\% of sun-like stars host planets with Msini = 1-3 earth masses. The similarity between the frequency of occurrence of exoplanets and the fraction of white dwarfs with metal lines within 25 pc is almost certainly just a coincidence since we do not know exactly how many of the white dwarfs in the 25 pc sample have been observed with high resolution.

Among the present 25 pc sample in Table 1, we have uncovered three white dwarfs with IR excesses from WISE photometry that were previously unknown. This is suggestive of debris disks (Cox,A., Sion, E., \& Debes,  J.2014) associated with the three white dwarfs.

\section{Preliminary Tests of Cool White Dwarf Spectral Evolution}

The total space density of white dwarfs within 25 pc remains unchanged from the 20 pc sample (Holberg et al. 2008) while the completeness drops from 85\% at 20pc to $\sim$60\% at 25 pc.
The space density of white dwarfs in Holberg et al. (2008) is derived from the virtual completeness of the white dwarfs within the core 13 pc sample. This can be determined in two ways, first from a direct count of white dwarfs within 13 pc, and by matching the slope of a log N - log (distance) plot to an expected slope of -3. The completeness estimates for the 20 pc and 25 pc samples are then simply computed with respect to the measured space density (see Holberg et al. 2013). Nevertheless by considering the 20 pc sample, the nature of the incompleteness at 65\% can be considered well characterized. We can use this knowledge to conduct a simple, preliminary, prospective test of white dwarf spectral evolution to test the hypotheses that the ordinary DQ stars (i.e. excluding the class of hot DQs) are descendants of cooling DB stars (Koester et al.1982; Wegner \& Yackovich 1984).

The formation rate of DB stars can be estimated 
from the following expression: 
$$
\frac{dN}{dt}   = f  \frac{N_{wd}}{ t_{DB} } 
$$
where f is the ratio of DB degenerates (Kleinman et al. 2004) to all other white dwarf types within the observed range of $M_{bol}$ for DB stars, $N_{wd}$ is the local space density of 
white dwarfs in the same range of $M_{bol}$ as the DB 
stars and $t_{DB}$ is the time spent by DB stars cooling
within their observed range of $M_{bol}$. $N_{wd}$ is 
obtained from a cool white dwarf luminosity function.

The temperature range of the DB white dwarfs
 is $30000K > T_{eff} > 12000K$ corresponding to a luminosity range 
$-1.01 > Log (L / L_{\odot}) > -2.52$. For DB stars, the local space 
density within the interval $12000K < T_{eff} < 30000K$,
$$
 N_{DB} \approx 2 \times 10^{-4} DBs/pc^3.
$$
and the parameter f = 20\% for DB stars (Kleinman et al. 2004). 
The 20 pc sample of Holberg et al. (2008) contained no spectroscopically 
identified DB stars.  Bergeron et al. (2011) have conducted a comprehensive 
analysis of 108 DB stars including spectroscopic distance estimates, 
finding four DBs within 25 pc.  One, WD2147+280, has a trigonometric 
parallax  (van Altena et al., 1994) which yields a distance of 
$35.7 \pm 0.4$ pc, which as the authors point out,  is difficult to 
reconcile with the estimated distance unless the spectroscopic gravity 
is reduced to 8.2 in which case it becomes consistent with the larger distance.  
After calculating distance uncertainties from the results of Bergeron 
et al. (2011), we have included three of these stars; WD1542-275, 
WD2058+392, and WD2316-173 in our new local sample including  appropriate 
distance uncertainties.  Overall, Bergeron et al. found a DB space density 
of $5.15 \times 10^{-5}~$pc$^{-3}$ which gives an expected number of 
approximately four DB white dwarfs in the 25 pc sample that compares 
favorably with three DB stars included in our sample.
If we assume an average DB mass,  
$M_{DB} =0.6 M_{\odot}$, then the cooling timescale for a DB from 30000K
to 12000K, $t_{DB} \approx 3.95 \times 10^8 $
years according to the cooling models of Bergeron et al.(2011 and references therein)\footnote{\url  http://www.astro.umontreal.ca/~bergeron/CoolingModels/}. Hence, 
the formation rate of DB stars, $dN_{DB} /dt$ is
$$
            \frac{dN_{DB}} { dt } =  1.01 \times 10^{-13} DB stars pc^-3 yr^-1
$$

Likewise for the formation rate of DQ stars 
from the following expression: 
$$
\frac{dN}{dt} = f  \frac{N_{wd }}{t_{DQ}} 
$$
where f is the ratio of DQ degenerates to all other
white dwarf types within the observed range of $M_{bol}$ 
for DQ stars, $N_{wd}$ is the local space density of
white dwarfs in the same range of $M_{bol}$ as the DQ 
stars and $t_{DQ}$ is the time spent by DQ stars cooling
within their observed range of $M_{bol}$. $N_{wd}$ is 
obtained from a cool white dwarf luminosity function.

The temperature range of the DQ stars is 
$12000K > T_{eff}  > 5500K$, corresponding to a luminosity range
$-2.52 > Log L / L_{\odot} > -3.88$. For the DQ stars
f = 9\%. Assuming that DQ masses are $M_{wd} = 0.6 M_{\odot}$, 
the cooling time for a DQ through its interval of $M_{bol}$ is $t_{DQ} = 3.6 \times 10^9$ years
The local space density of DQs within the 
Interval $12000K > T_{eff} > 5500K$, $N_{DQ} \approx  2 \times 10^{-3} DQs/pc^3$.
Thus, the formation rate of DQ stars, $dN_{DQ} /dt$
is $dN_{DQ} / dt  =  4.5 \times 10^{-14}  pc^{-3} /yr$. 
The lower formation for the DQ stars relative to DBs presumably may reflect that not all cooling DB stars become DQ stars and may become DA stars either through accretion of volatile-rich tidally disrupted bodies, interstellar H, or diffusive float up of H 
as the DB cools. This very tentative result must be confirmed with larger, more complete samples of white dwarfs.

Although this evolution test is only preliminary (given the uncertainties in the parameters used), it is nonetheless illustrative of the far-reaching potential of enlarging the volume of space around the sun containing known white dwarfs, thus increasing the sample size of
of white dwarfs in each astrophysically important spectroscopic subclass including the magnetic degenerates as well as the DA, DB, DQ, DZ and DC degenerates. 

\section{Conclusions}

Our study has revealed the distribution of white dwarf spectral types for 224 degenerate stars in a volume-limited sample out to 25 parsecs. We find the following characteristics of the sample:

(1) Little or no evidence of halo or thick disk component members within 25 pc but seven degenerates with extraordinarily high space motions; 

(2) The sample includes a sizable number of DQ stars and cool (mostly DC and DZ) white dwarfs. We note a possibly significant pileup of DC stars at the low temperature cutoff of the DQ stars;

(3) The incidence of magnetic white dwarfs within
25 pc is 8\% of the total sample. This is close to the estimate of the true incidence of field white dwarfs in the galactic disk by Liebert et al.(2003);

(4) We carry out a preliminary test of one scenario of white dwarf spectral evolution theory, namely that all of the DQ stars
(excluding the "hot" DQs) are the evolutionary descendants of the DB white dwarfs. Within the uncertainties in the true space densities of DB stars and DQ stars, we find preliminary evidence that the formation rate of DQ stars is smaller than the DB rate, suggesting that not all DB stars evolve into DQ stars below 12,000K; 

(5) We find no compelling evidence of any significant differences between the space motions of the various spectroscopic subtypes.

As this study indicates the local WD population, with its high degree of completeness, provides an invaluable ground truth sampling of the WD population, particularly in the galactic disk.  As such it affords a number of ways to help characterize much larger, deeper and more distant WD samples that are not volume limited, for example those from the Sloan Digital Sky Survey, (SDSS).  Towards that end, compelling motivation now exists to both increase the completeness of the current 25 pc sample and extend it to greater distance, thus providing  a larger basic sample.  Such efforts will also set the stage for a time during the next decade when the European Space Agency Gaia mission will make possible a virtually complete determination of all WD  distances and proper motions (but not necessarily stellar spectra) out to $\sim$100 pc. Also within the same time frame large surveys such as the Large Synoptic Survey Telescope (LSST) will be sampling WD populations well into the halo.

This work is supported by NSF grant 1008845. We are grateful for several helpful comments and corrections from an anonymous referee. We thank Detlev Koester for insightful comments on the spectra of cool DQ stars. We thank John Debes for useful input on the frequency of exoplanets around sun-like stars. J.B.H  also wishes to acknowledge NASA Astrophysics Data Program grant NNX1OAD76.  This research has made use of the White Dwarf Catalog maintained at Villanova University and the SIMBAD database, operated at CDS, Strasbourg, France.

{\bf{REFERENCES}} 

Aannestad, P. et al.1993, AJ, 105, 1033

Bergeron, B., et al. 2011, ApJ, 737, 28

Bergeron, P., Leggett, S., \& Ruiz, M.-T.2001, ApJS, 133, 413

Chiba, M.,\& Beers, T. 2000, AJ, 119, 2843

Cox,A., Sion, E., \& Debes,  J.2014, BAAS, in press

Dufour, P., Bergeron, P., \& Fontaine, G. 2005, ApJ, 627, 404

Dufour, P., Bergeron, P., Liebert, J., Harris, H., Knapp, G., Anderson, S., Hall, P., Staruss, M., Collinge, M., \& Edwards, M. 2007,ApJ, 663, 1291

Farihi, J., Jura, M., \& Zuckerman, B. 2009, ApJ, 694, 805

Giammichele, N., Bergeron, P., Dufour, P.2012, ApJS, 199, 29
 
Gianninas, A., Bergeron, P., \& Ruiz, M.-T. 2011, ApJ, 743, 138

Gliese, W.,\& Jarheise, H.1995, Third Catalog of Neartby Stars,  

Greenstein, J. 1986, ApJ, 304, 334

Greenstein, J., \& Trimble, V.1967, ApJ, 149, 283

Hall, P.B., \& Maxwell, A.J.2008, ApJ, 678, 1292

Holberg, J. B. 2008, Talk, 16th European White Dwarf Workshop (in press)

Holberg, J. B., Sion, E. M., Oswalt, T, McCook, G. P., Foran, S., \& Subasavage, J. P. 2008, AJ 135, 1225 (LS08) 

Holberg, J. B., Bergeron, P. \& Gianninas, A. 2008, AJ, 135, 1239

Holberg, J. B., Bergeron, P. 2006, AJ, 132, 1221

Holberg, J., et al.2013, in preparation

Howard,A.2013, Science, 340, 572

Kawka, A. \& Vennes, S. 2006, ApJ, 643, 402

Kilic, M., et al.2010, ApJS, 190, 77

Kleinman, S.J., et al.2004, ApJ, 607, 426 

Koester, D., Weidemann, V., \& Zeidler,E.-M. 1982, A\&A, 116, 147

Koester, D., \& Weidemann, V., et al., A\&A, 530, 114

Koester, D.; Girven, J.; Gänsicke, B. T.;  Dufour, P.2011,  A\&A,  530, 114 

Koester, D., \& Knist, S.2006, A\&A 454, 951

Liebert, J, Dahn, C,. \& Monet, D.1988, ApJ, 332, 891

Liebert, J., Bergeron, P., \& Holberg, J.B.2003, AJ, 125, 348

Limoges, M.-M., Lepine, S., \& Bergeron, P.2013, AJ, in press (arXiv:1303.2094)

Lepine, S. et al. 2009, AJ, 137, 4109

Maxted,P., Marsh, T.,\& Moran, C.2000, MNRAS, 319, 305  

Mayor, M., Marmier, C., Lovis, S., Udry, D., S´egransan, F., Pepe, Benz, W., Bertaux,J.-L., Bouchy,F., 
Dumusque, X., LoCurto, G., Mordasini, C., Queloz, D., and Santos
N.C.

McCook, G.P., \& Sion, E.M. 1999, ApJS, 121, 1

Oswalt, T.D.1994, private communication.

Oswalt, T., et al. 1996, Nature, 382, 692

Pauli, E.-M. et al. 2003, A\&A, 400, 877

Pauli, E.-M.2006, A\&A, 447, 173

Saffer, R., et al. 1993, AJ, 105, 1945

Sayres, C., Subasavage, J. P., Bergeron,
P., Dufour, P., Davenport, J. R. A., AlSayyar, Y., \& Tofflemire, B. M.,  2012, AJ, 143, 103

Silvestri, N. et al. 2001, AJ, 121, 503

Silvestri, N.,  Oswalt, T. D., \& Hawley,  S.2002, AJ, 124, 1118 

Sion, E. 1984, ApJ, 282, 612 

Sion, E. M., Fritz, M. L., McMullin, J.  P., \& Lallo, M. D. 1988, AJ, 96, 251

Sion, E.M., Holberg, J.B., Oswalt, T.D., McCook, G.P., \& Wasatonic, R.2009, AJ, 138, 1681

Subasavage, J., Henry, Todd J., Bergeron, P., Dufour, P., Hambly, N., Beaulieu, T.D. 2007, AJ, 134, 252 

Subasavage J.P., Henry, T.J., Bergeron P. Dufour, P. \& Hambly, N. C. 2008, AJ, 136, 899

Subasavage, J. et al. 2009, AJ, 137, 4547

Trimble, V.,\& Greenstein, J.L.1972, ApJ, 179, 441

Tremblay, P.-E., \& Bergeron, P. 2008,  ApJ,672, 1144

Tremblay, P.-E.; Bergeron, P.; Gianninas, A.2011, ApJ, 730, 128 

Wegner,G.1974, MNRAS, 166, 271

Wegner, G., \& Yackovich, F. H. 1984, ApJ, 284, 257

Wegner, G., \& Reid, N.1987, in Faint Blue Stars,IAU Colloq.\#95 

Weidemann, V., \& Koester, D.1989, A\&A, 210, 311

Wooley, R., et al. 1970, Ann. R. Obs., 5

Zacharias, N., Finch, C.T., Girard, T.M., Henden, A., Bartlett, J.L., Monet, D.G., \&
Zacharias, M.I. 2012, UCAC4 Catalog, VizieR On-line Data Catalog: I/322.

\clearpage


\begin{figure}
\plotone{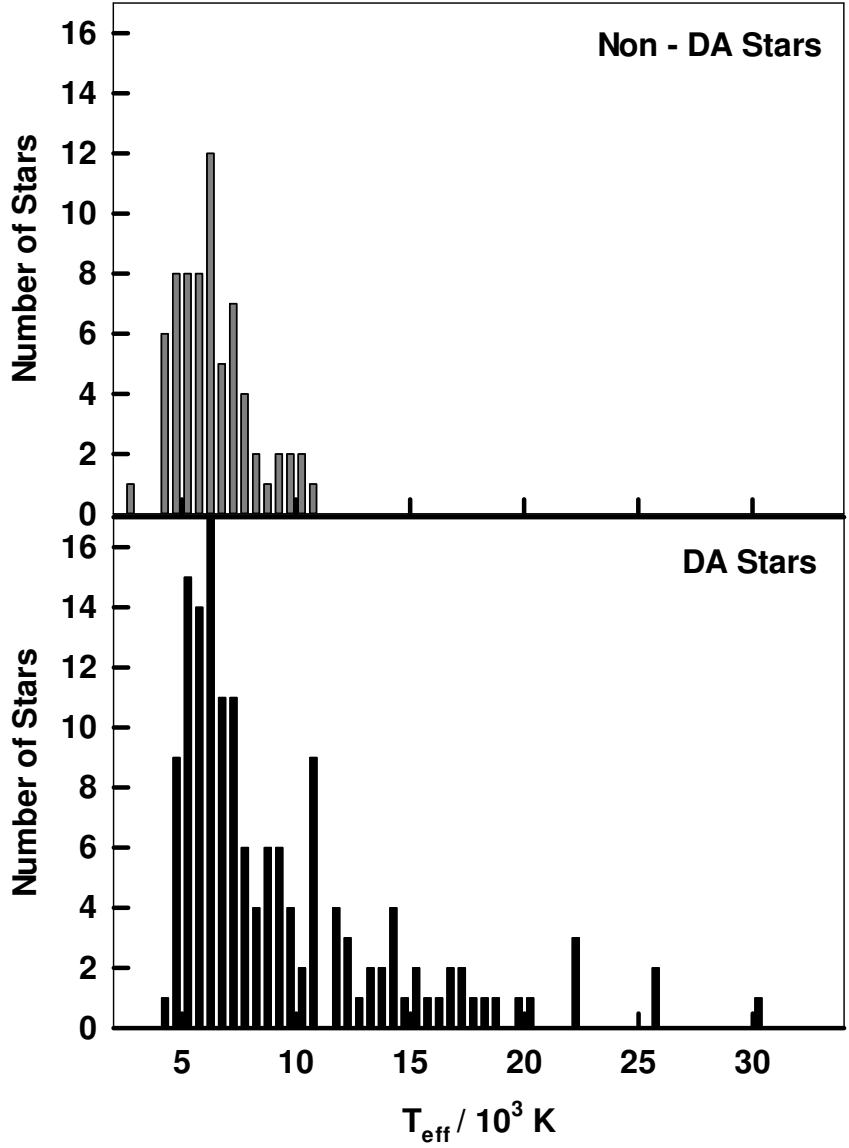}
\caption{Histogram of the number versus surface temperature distribution function for all DA white dwarfs (lower panel) and all non-DA white dwarfs (upper panel) within 25 pc of the sun}
\end{figure}

\begin{figure}

\plotone{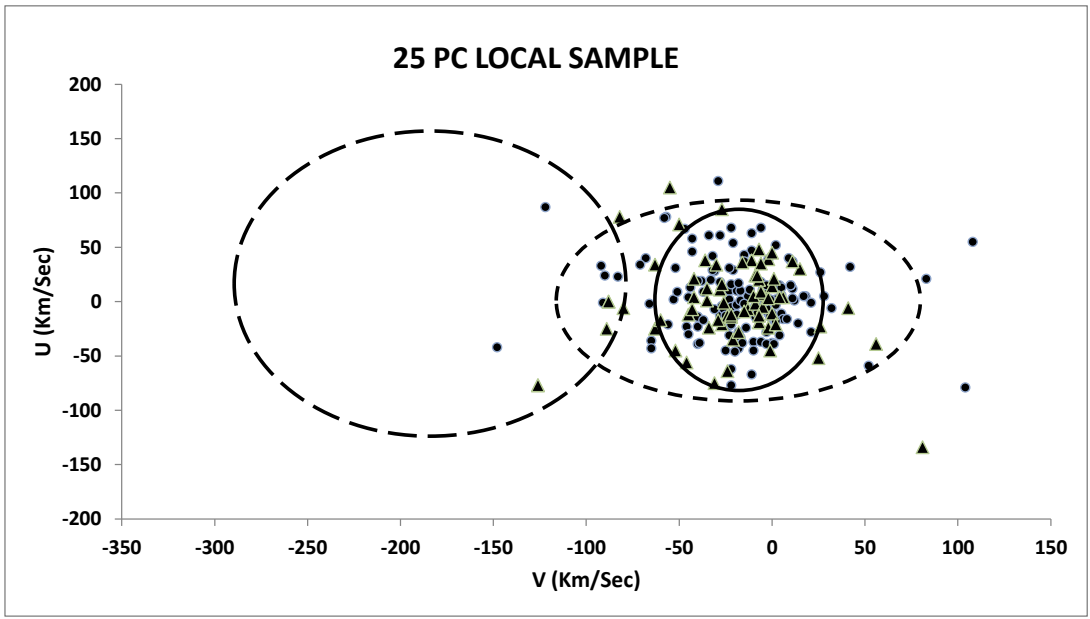}

\caption{The U versus V space velocity diagram for the 25 pc sample of white dwarfs with the assumption of zero radial velocity. DA stars are denoted by closed circles, while non-DA stars are denoted by closed triangles. For comparison, three velocity ellipses for main sequence stars are shown following Chiba \& Beers (2000; see also Vennes \& Kawka 2006),
the 2 $\sigma$ velocity ellipse contour (solid line) of the thin disk component, the 2 $\sigma$ ellipse of the
thick disk component (short-dashed line) and the 1 $\sigma$ contour 
of the halo component (long-dashed line).}

\end{figure}

\clearpage

\begin{figure}

\plotone{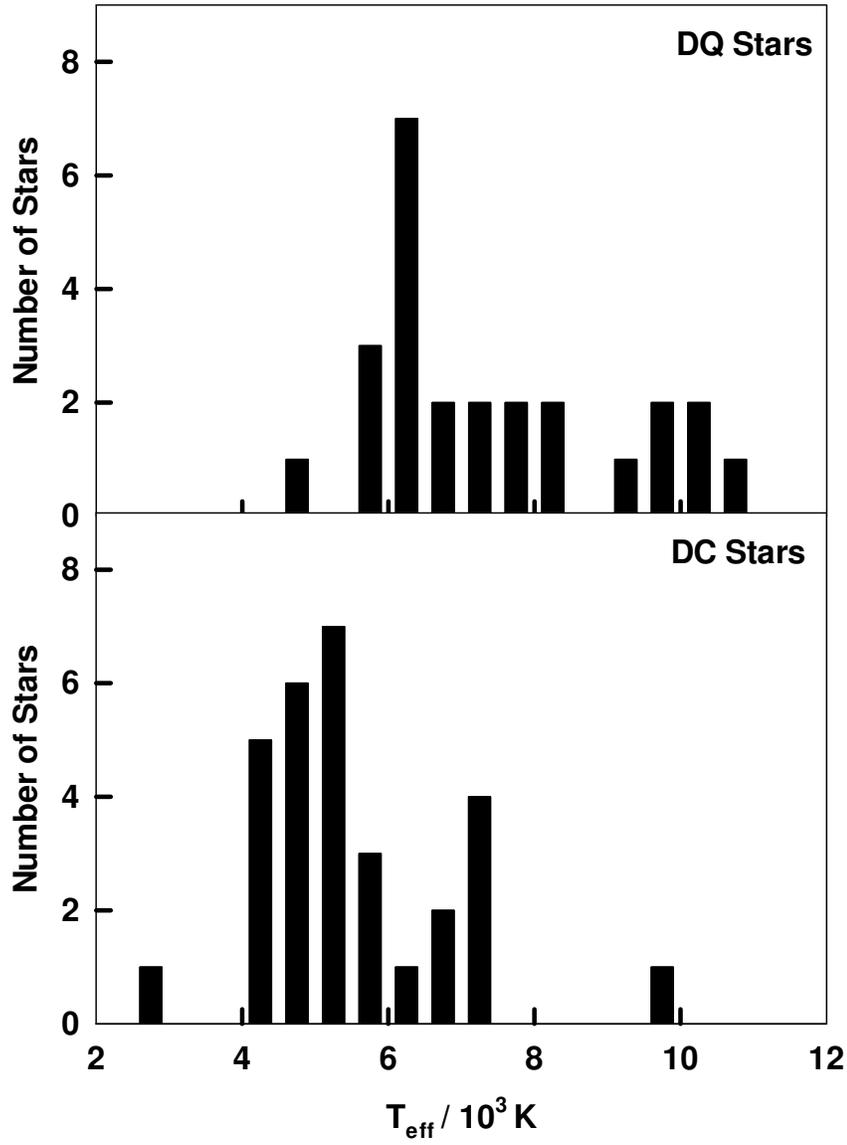}

\caption{Histogram of the number versus surface temperature 
distribution function for all DQ white dwarfs (upper panel) 
and all DC white dwarfs (lower panel) within 25 pc of the sun. 
Note the pileup of DC stars near the temperature at which the DQ stars 
show a real cutoff.}

\end{figure}

\clearpage

\setlength{\hoffset}{-20mm} 
\begin{deluxetable}{lccccccccc}
\tablecaption{Observational Data Used in Space Motions}
\tablewidth{0pc}
\tablenum{1}
\tablehead{ 
\colhead{WD}
   &\colhead{TYPE}
            &\colhead{$T_{eff}$}
                           &\colhead{RA}
                                      &\colhead{DEC}           
                                                &\colhead{DIST}
                                                        &\colhead{PM}
                                                                 &\colhead{PA}
                                                                         &\colhead{method}
&\colhead{Ref} }
\startdata 

0000-345  & DCP8.1  & 6643 &  000.667 & -34.222 & 13.21 & 0.7578 & 169.046 & p & L20 \\

0008+424   & DA6.8  & 7380 &  002.843 & +42.678 & 22.00 & 0.2328 & 191.648 & sp & L20\\

0009+501   & DAH7.6  & 6502 &  003.061 & +50.422 & 11.03 & 0.7150 & 219.920 & p & L20 \\

0011-134   & DAH8.4  &5992 &  003.553 & -13.183 & 19.49 & 0.8990 & 217.337 & p& L20 \\

0011-721   & DA7.8  & 6325   & 003.457 & -71.831 &   -    & 0.3260 & 141.300 & sp & GBD \\

0029-031   & DA11.3  & 4470 & 008.041  & -02.900 & 23.47 & 0.6505 &  76.305 & p & pi  \\

0038+555   & DQ4.6 & 10900   & 010.337 & +55.834 & 23.04 & 0.3396 & 103.627 & p& pi \\

0038-226   & DQpec9.3 & 5529  & 010.358 & -22.350 & 9.04  & 0.6047 & 232.600 & p & L20\\

0046+051   & DZ7.4 & 6215   & 012.291 & +05.388 & 04.31 & 2.9780 & 155.538 & p & L20\\

0053-117  &  DA7.1 &        & 13.95975 & -11.45875 & -  & 0.4396 & 350.045 & sp & G11 \\

0108+048   & DA6.4 & 8530   & 015.958 & +05.075 & 21.32 & 0.3924 & 053.393 & p & G11\\

0108+277   & DA9.6 & 6428  & 017.686 & +27.970 &   13.8    & 0.2270 & 219.321 & p & L20 \\

0115+159   & DQ5.6 & 9119  & 019.500 & +16.172 & 15.40 & 0.6480 & 181.805 & p & L20\\

0121-429   & DAH7.9 & 6299  & 021.016 & -42.677 & 18.31 & 0.5941 & 151.000 & p &L20 \\

0123-262   & DC6.9 &        & 21.35194 &-26.01238 &-    & 0.5945 & 154.8143 & sp & GBD\\

0123-460   & DA8.5& 5898   & 021.325 & -45.752138 & 24.90 & 0.7483 & 136.299 & sp & Sub08\\

0134+883   & DA2.8 & 18311  & 025.369 & +83.583 & 25.60 & 0.1500 & 306.870 & sp & G11\\

0135-052   & DA6.9 & 7118   & 024.497 & -04.995 & 12.34 & 0.6810 & 120.838 & p & L20 \\

0141-675   & DA7.8 & 6248  & 025.754 & -67.308 & 09.72 & 1.0797 & 199.000 & p & L20\\

0145+360   &   DA7.8 & 6470  &   27.168 & +36.258  &   24.2  &     -  & -     &   sp  & Lim \\

0148+467   & DA3.8 & 14005   & 028.012 & +47.001 & 15.85 & 0.1240 &   0.569 & p & L20 \\

0148+641   & DA5.6 & 9016  & 027.963 & +64.431 &   17    & 0.2854 & 123.857 & sp & DR7 \\

0208+396   & DAZ6.9 &7264 & 032.836 & +39.922 & 16.72 & 1.1450 & 115.746 & p & L20\\

0210-508   & DQ+K1 & 6000  & 032.608 & -50.823 & 10.91 & 2.1920 &  72.666 & p  & SL\\

0213+396   & DA5.4 &       & 34.06816 & 39.85708 &-    & 0.1873 & 239.876 & sp & G11\\

0213+427   & DA9.0 &5507   & 034.281 & +42.977 & 19.92 & 1.0470 & 125.065 & p &n L20\\

0227+050   & DA2.7& 18779   & 037.569 & +05.264 & 24.30 & 0.0783 & 107.764 & p & G11\\

0230-144   & DA9.2 & 5477  & 038.157 & -14.197 & 15.62 & 0.6870 & 177.114 & p& L20 \\

0231-054   & DA3.7 & 13550   & 038.532 & -05.194 & 22.7  & 0.2655 &  69.722 & sp& GBR11 \\

0233-242   & DC9.3 & 5312  & 038.840 & -24.013 &   15.3    & 0.6220 & 189.015 & sp & L20\\

0236+259   & DA9.2 & 5500  & 039.832 & +26.165 & 21.00 & 0.3591 & 117.35  & p& Pi \\

0243-026   & DAZ7.4& 6839  & 041.628 & -02.456 & 21.23 & 0.5342 & 155.21  & p  & G11\\

0245+541   & DAZ9.5 & 5319  & 042.151 & +54.389 & 10.35 & 0.5735 & 227.827 & p & L20\\

0252+497   & DA7.9 & 6370   &    -     &   -      &  16.7  &   -   &   - & sp & Lim\\

0255-705   & DAZ4.7& 10560  & 044.071 & -70.369 & 21    & 0.6596 &  99.776 & sp & G11\\

0310-688   & DA3.3 & 16865  & 047.628 & -68.600 & 10.15 & 0.1112 & 158.097 & p & L20\\

0311-649   & DA4.0 &11945  & 048.107 & -64.736 & 21.00 & 0.1661 & 101.107 & sp & Sub08\\

0322-019   & DAZ9.9 & 5195 & 051.296 & -01.820 & 16.80 & 0.9090 & 164.625 & p & L20\\

0326-273   & DA5.4 & 8483 & 052.203 & -27.317 & 17.36 & 0.8503 & 071.629 & p & L20\\

0340+198   &  DA7.0 & 7160     &   55.84638      &   19.97049     & 18.4  &  0.178 &  24.55 & sp & Lim\\

0341+182   & DQ7.7 &6568  & 056.145 & +18.436 & 19.01 & 1.1990 & 159.771 & p& L20 \\

0344+014   & DC9.9 & 5170   & 056.778 & +01.646 &  -       & 0.4730 & 150.400 & sp & L20 \\

0357+081   & DA9.2 &5478  & 060.111 & +08.235 & 17.82 & 0.5352 & 222.274 & p & L20\\

0413-077   & DA3.1 & 17100   & 063.839 & -07.656 & 05.04 & 4.0880 & 213.216 & p & L20\\

0416-594   & DA3.3 &14000   & 064.122 & -59.302 & 18.23 & 0.1740 & 195.838 & p & SL \\

0419-487   & DA8 & 6300    & 065.273 & -48.652 & 20.12 & 0.5402 & 178.302 & p  & G11\\

0423+044   & DA &5140    & 066.658 & +04.541 & 20.72 & 0.8469 & 131.936 & p& Pi \\

0423+120   & DA8.2 & 6167  & 066.473 & +12.196 & 17.36 & 0.2446 & 335.866 & p & L20\\

0426+588   & DC7.1 &7178  & 067.797 & +58.977 & 05.53 & 2.4267 & 147.602 & p & L20\\

0431-279   & DC9.5  & 5330 & 068.390 & -27.890 & 24.7  & 0.388  &  90.738 & p & Sub08\\

0431-360   & DA10.0 &5153  & 068.232 & -35.395 & 25.0  & 0.301  &  84.1   & sp & Sub08\\

0433+270   & DA9.0 &5629  & 069.187 & +27.164 & 17.85 & 0.2760 & 124.196 & p & L20 \\

0435-088   & DQ8.0 &6367  & 069.447 & -08.819 & 09.50 & 1.5740 & 171.103 & p & L20\\

0454+620   & DA4.3 & 11610   & 24.663 & +62.152 & 24.9 &   0.188  & 149.6 & sp & Lim\\

0457-004   & DA4.7 & 10800  & 074.930 & -00.377 & 21.93 & 0.2926 & 142.778 & sp& L20 \\

0503-174   & DAH9.5& 5300  & 076.498 & -17.378 & 20.27 & 0.6876 & 14.4    & p & Pi\\

0511+079   & DA7.7 & 6590  & 078.514 & +08.004 & 20.26 & 0.3642 & 216.178 & p& G11 \\

0532+414   & DA6.8 &7739  & 084.084 & +41.498 & 23.81 & 0.1668 & 284.903 & sp& G11 \\

0548-001   & DQP8.3 & 6070 & 087.831 & -00.172 & 11.07 & 0.2510 & 025.810 & p & L20\\

0552-041   & DZ10.0 & 5182  & 088.789 & -04.168 & 06.41 & 2.3760 & 166.600 & p \\

0553+053   & DAP8.7& 5785  & 089.106 & +05.363 & 07.99 & 1.0272 & 204.993 & p & L20\\

0615-591   & DB3.2 & 16714   & 094.063 & -59.206 & 23.80 & 0.379  & 166     & p & Pi\\

0618+067   & DA8.1 & 5940  & 095.198 & +06.754 & 22.62 & 0.5352 &  91.392 & p & G11\\

0620-402   & DZ6 &5919    & 095.423 & -40.217 & 21.50 & 0.379  & 166     & p & L20\\

0628-020   & DA  &6912    & 097.661 & -02.097 & 21.50 & 0.2088 & 253.300 & p & G11\\

0642-166   & DA2 &25967    & 101.288 & -16.713 & 02.63 & 1.3394 & 204.057 & p& L20 \\

0644+025   & DA6.8 &22288  & 101.842 & +02.519 & 18.45 & 0.4234 & 272.572 & p & L20\\

0644+375   & DA2.4 &22288  & 101.908 & +37.515 & 15.40 & 0.9624 & 193.561 & p & L20\\

0649+639   & DA8.1 & 6230   &  103.560 & +63.932 & 21.1 &   0.200  & 225.0 & sp & Lim\\

0651-398A  & DA7.0 & 7222  & 103.397 & -39.925 & 25.10 & 0.2125 & 341.906 & sp  & Sub08\\

0655-390   & DA7.9 &6311   & 104.274 & -39.159 &   17.2    & 0.3400 & 242.600 & sp & GBD\\

0657+320   & DA10.1 & 4888  & 105.215 & +31.962 & 15.19 & 0.6910 & 149.362 & p & L20\\

0659-063   & DA7.7 &6627   & 105.478 & 
-06.463 & 12.34 & 0.8984 & 184.981 & p & L20\\

0706+377   & DQ7.6 &6590   & 107.559 & +37.672 & 24.27 & 0.3577 & 220.010 & p& Pi \\

0708-670   & DC9.9 &5097  & 107.217 & -67.108 &   17.5    & 0.2460 & 246.300 & sp & GBD \\

0727+482.1 & DA10.0 &4934 & 112.678 & +48.199 & 11.11 & 1.2868 & 190.070 & p & L20\\

0727+482.2 & DA10.1 &4926 & 112.697 & +48.173 & 11.11 & 1.2868 & 190.070 & p & L20 \\

0728+642   & DAP11.1& 5135 & 113.378 & +64.157 &   13.4    & 0.2660 & 171.352 & sp & L20\\

0736+053   & DQZ6.5& 7871  & 114.827 & +05.227 & 03.50 & 1.2590 & 214.574 & p& L20 \\

0738-172   & DZA6.6  & 7650   & 115.086 & -17.413 & 09.09 & 1.2634 & 116.600 & p & L20\\

0743-336   & DC10.6 & 4462  & 116.410 & -33.931 & 15.19 & 1.7360 & 352.670 & p & L20 \\

0744+112   &  DA6.2  & 8160 & 116.87546 & 11.12626  & 25.7  &  0.1907      &  161.66   & sp & Lim\\

0747+073.1 & DC10.4 &4366 & 117.563 & +07.193 & 18.28 & 1.8049 & 173.414 & p & L20 \\

0747+073.2 & DC12.0 & 4782  & 117.560 & +07.196 & 18.28 & 1.8049 & 173.414 & p & L20\\

0749+426   & DC11.7 &4585 & 118.305 & +42.500 &   17.8    & 0.420  & 165.845 & sp & L20 \\

0751-252   & DA9.8 &5085  & 118.485 & -25.400 & 17.68 & 0.3622 & 304.700 & p & L20\\

0752-676   & DA8.8 &5735  & 118.284 & -67.792 & 07.89 & 2.1499 & 135.867 & p & L20\\

0753+417   & DA7.3 & 6880   & 119.129 & +41.664 & 24    & 0.3481 & 181.316 & sp  & Kilic-10\\

0805+356   & DA7.3 &6900  & 122.296 & +35.465 & 21.1  & 0.1051 & 220.873 & sp & Trem11\\

0806-661   & DQ4.9 &10205  & 121.723 & -66.304 & 19.16 & 0.4468 & 130.400 & p & L20\\

0810+489   & DC6.9 &7300   & 123.546 & +48.758 & 17.00 & 0.258  & 166.5   & p & GBD \\ 

0816-310   & DZ7.6 & 6463  & 124.667 & -31.172 & 23.80 & 0.8163 & 164.074 & sp & GBD \\

0821-669   & DA9.8 &5088  & 125.361 & -67.055 & 10.65 & 0.7623 & 329.500 & p& L20 \\

0827+328   & DA6.9 &7490   & 127.664 & +32.696 & 22.27 & 0.5412 & 196.090 & p & L20\\

0839-327   & DA5.5 & 9081  & 130.384 & -32.942 & 08.81 & 1.7020 & 322.700 & p & G11 \\

0840-136   & DZ10.3 & 4874 & 130.701 & -13.786 &    19.3   & 0.2720 & 263.000 & sp & L20 \\

0843+358   & DZ6 & 9041    & 131.685 & +35.642 & 23.1  & 0.174  & 244.466 & sp  & GBD\\

0856+331   & DQ5.1 &9920  & 134.811 & +32.953 & 20.49 & 0.334  & 269.657 & p & Pi\\

0912+536   & DCP7 &7235   & 138.983 & +53.423 & 10.30 & 1.5630 & 223.997 & p& L20 \\

0946+534   & DQ6.2 & 8100  & 147.571 & +53.254 & 22.98 & 0.2642 & 259.753 & p & Pi \\

0955+247   & DA5.8 &8621  & 149.451 & +24.548 & 24.44 & 0.4200 & 219.848 & p & L20 \\

0959+149   & DC7 &7200    & 150.455 & +14.689 & 22.2  & 0.339  & 269.493 & p & GJ \\

1008+290   & DQpec11.0 &   & 152.92329 & 28.76649 & -   & 0.7201 & 189.593 & p& Pi\\

1009-184   & DZ8.5 & 6036  & 153.007 & -18.725 & 18.30 & 0.5114 & 269.000 & p & L20\\

1012+083.1 & DA7.5& 6750   & 153.760 & +08.109 & 25.80 & 0.3339 & 300.405 & sp & G11\\

1019+637   & DA7.2 &6742  & 155.787 & +63.461 & 16.33 & 0.3790 & 053.160 & p  & L20\\

1033+714   & DC10.3 &4727 & 159.260 & +71.183 & 15.15 & 1.9170 & 256.008 & p & L20\\

1036-204   & DQpecP10.2 &4694 & 159.731 & -20.682 & 14.28 & 0.6100 & 334.000 & p & L20 \\

1043-188   & DQ8.1 &5780  & 161.412 & -19.114 & 12.15 & 1.9780 & 251.636 & p & L20\\

1055-072   & DA6.8 &7491   & 164.396 & -07.523 & 13.42 & 0.8270 & 276.328 & p & L20\\

1105-048   & DA3.5 &15141  & 166.999 & -05.157 & 24.2  & 0.4445 & 188.148 & sp & G11\\

1116-470   & DC8.6 &5801  & 169.613 & -47.365 & 17.9  & 0.3220 & 275.100 & sp & GBD\\

1121+216   & DA6.7 &7434  & 171.054 & +21.359 & 13.42 & 1.0400 & 269.240 & p & L20\\

1124+595   & DA4.8&10747   & 171.718 & +59.321 & 25    & 0.1569 & 108.204 & sp & L20\\

1132-325   & DC &     & 173.623 & -32.832 & 09.53 & 0.9400 & 038.955 & p & L20\\

1134+300   & DA2.4 & 22469   & 174.271 & +29.799 & 15.31 & 0.1480 & 267.948 & p & L20 \\

1142-645   & DQ6.4 &7966  & 176.458 & -64.841 & 04.62 & 2.6876 & 097.500 & p & L20\\

1148+687   & DA7.6 &   - &   177.71808 & +68.521 &  
16.8 & - & - & sp & DR7\\

1149-272   & DQ8.1 & 6200  & 177.900 & -27.539 & 24.40 & 0.229  & 283.126 & sp  & Sub08\\

1202-232   & DAZ5.8 & 8767 & 181.361 & -23.553 & 10.82 & 0.2458 &  16.600 & p & L20 \\

1208+576   & DAZ8.6 & 6200 & 182.872 & +57.404 & 20.44 & 0.5486 & 132.414 & p & G11 \\

1214+032   & DA8.0 & 6272  & 184.216 & +02.968 & 19.72 & 0.6947 & 291.357 & sp & Pi \\

1223-659   & DA6.6 & 7594   & 186.625 & -66.205 & 16.25 & 0.1858 & 186.900 & p & G11\\

1236-495   & DA4.3 & 11599  & 189.708 & -49.800 & 16.39 & 0.4902 & 255.709 & p & L20 \\

1241-798   & DC/DQ & 9556   & 191.219 & -80.157 & 22.10 & 0.5524 & 309.858 & sp & Sub08\\

1242-105   & DA6.3 &        & 191.21941 &-10.85241 &-& 0.3488 & 257.079 & sp & GBD\\

1257+037   & DA9.0& 5616   & 195.037 & +03.478 & 16.58 & 0.9696 & 206.195 & p & L20 \\

1 309+853   & DAP9 & 5440  & 197.171 & +85.041 & 16.47 & 0.3213 & 140.811 & p & L20 \\

1310+583   & DA4.8 & 10544   & 198.241 & +58.086 & 21.1  & 0.1995 & 112.708 & sp & G11 \\

1310-472   & DC11.9 &4158  & 198.248 & -47.468 & 15.03 & 2.2047 & 105.253 & p & L20 \\

1315-781   & DC8.8 & 5619  & 199.856 & -78.391 & 19.17 & 0.4700 & 139.5   & p& Pi \\

1327-083   & DA3.6 & 14571  & 202.556 & -08.574 & 16.86 & 1.2049 & 246.761 & p & L20 \\

1334+039   & DA11 & 4971   & 204.132 & +03.679 & 08.23 &  3.8800 & 252.775 & p & L20\\

1344+106   & DAH7.1 & 7059     & 206.851 & +10.360 & 20.04 & 0.9032 & 260.569 & p & L20 \\

1337+705   & DAZ2.5 & 20464 & 204.710 & +70.285 & 24.79 & 0.405  & 266.035 & p &G11 \\

1338+052   & DC11.6 &       & 205.34083 & 5.01272 &- & 0.438 & 271.6 & sp & Say \\

1339-340   & DA9.5 & 5361  & 206.508 & +57.009 & 21.20 & 0.2758 & 315.881 & sp & GBD\\

1344+572   & DA3.8 & 13389  & 206.510 & +57.008 & 20    & 0.273  & 315.9   & sp & G11 \\

1345+238   & DA11 & 4581  & 207.012 & +23.579 & 12.06 & 1.4960 & 274.637 & p & L20\\

1350-090   & DAP5 & 9518   & 208.314 & -09.275 & 22.73 & 0.3618 & 174.226 & sp & G11 \\

1401+457   & DC19 & 2600   & 210.853 & +45.558 & 24.00 & 0.2840 & 251.947 & sp & Kilic$_{10}$ \\

1425-811   & DAV4.2 & 12098 & 218.282 & -80.157 & 22.72 & 0.5524 & 201.418 & p & G11\\

1436-781   & DA8.1 & 6270  & 220.714 & -78.398 & 24.65 & 0.4096 & 275.1   & p & Pi \\

1444-174   & DC10.2 & 4982  & 221.855 & -17.704 & 14.49 & 1.1440 & 252.643 & p & L20\\

1532+129   & DZ6.7 & 7500  & 233.774 & +12.795 & 22    & 0.2464 & 222.039 & sp & Koe$_{11}$\\

1538+333   & DA5.6 & 8940  & 235.139 & +33.147 & 22.72 & 0.186  & 296.152 & p & Pi\\

1542-275   & DB4.0 &       &  237.65833 & 27.65611 &-& 0.246 & 235.9 & sp & Ber11\\

1544-377   & DA4.8 & 10610  & 236.875 & -37.918 & 15.24 & 0.4685 & 242.838 & p & L20\\

1609+135   & DA5.4 & 9041  & 242.856 & +13.371 & 18.34 & 0.5510 & 178.513 & p & L20 \\

1620-391   & DA2.1  & 25985  & 245.890 & -39.229 & 12.86 & 0.0755 &  89.962 & p & L20 \\

1625+093   & DA7.3 & 7038  & 246.972 & +09.204 & 23.36 & 0.4872 & 192.325 & p & G11\\

1626+368   & DZA6.0 & 8507 & 247.104 & +36.771 & 15.94 & 0.8881 & 326.668 &  p & L20 \\

1630+089   & DA9.0 & 5640     & 177.717 & +68.521 &  13.2 &    0.4 &   -   & sp& Lim \\

1632+177   & DAZ5.0 & 10225  & 248.674 & +17.609 &   -    & 0.0885 & 108.435 & sp & L20 \\

1633+433   & DAZ7.7& 6608  & 248.755 & +43.293 & 15.10 & 0.3730 & 144.151 & p & L20\\

1633+572   & DQ8.2 & 5958  & 248.589 & +57.169 & 14.45 & 1.6440 & 317.229 & p & L20 \\

1639+537   & DAH6.7& 7510  & 250.238 & +53.685 & 21.09 & 0.2369 & 212.416 & p & Pi\\

1647+591   & DAV4.1 & 12738 & 252.106 & +59.056 & 10.95 & 0.3236 & 154.498 & p & L20\\

1655+215   & DAB5.4 & 9179  & 254.291 & +21.446 & 23.25 & 0.5820 & 178.040 & p & L20\\

1658+440   & DAP1.7 & 30510 & 254.951 & +44.017 & 22.00 & 0.1012 & 341.869 & sp & G11\\

1705+030   & DZ7.7& 6584   & 257.033 & +02.960 & 17.54 & 0.3790 & 180.907 & p & L20\\

1748+708   & DQ9.0 & 5570  & 267.033 & +70.876 & 06.07 & 1.6810 & 311.394 & p & L20\\

1756+143   & DA9.0 & 5466  & 269.595 & +14.293 & 22.40 & 1.0054 & 235.045 & sp & GBD\\

1756+827   & DA6.9 & 7214  & 267.458 & +82.773 & 15.64 & 3.5897 & 336.542 & p& L20 \\

1814+134   & DA9.5 & 5251   & 274.277 & +13.473 & 14.22 & 1.2070 & 201.500 & p& L20 \\

1817-598   & DA5.8 & 4960  & 275.497 & -59.863 & 24.87 & 0.3653 & 194.911 & p & Pi\\

1820+609   & DA10.5 & 4919 & 275.332 & +61.018 & 12.78 & 0.7133 & 168.517 & p & L20\\

1829+547   & DQP8.0 & 6345  & 277.584 & +54.790 & 14.97 & 0.3991 & 317.234 & p & L20\\

1840+042   & DA5.8 & 9090  & 280.857 & +04.339 & 24.87 & 0.187  & 296.935 & p & G11 \\

1900+705   & DAP4.2 & 11835 & 285.042 & +70.664 & 12.98 & 0.5064 &  10.467 & p & L20\\

1911+536   & DA2.9 & 17670    & 288.202 & +53.720 & 22.1 &   0.018  &   225.0   & sp & Lim\\

1912+143   & DA7.3 & 6940   & 288.650 & +14.473 & 19.4 & 0.161      & 225.0   &  sp & Lim\\

1917+386   & DC7.9 &6459  & 289.744 & +38.722 & 11.69 & 0.2510 & 174.028 & p & L20\\

1917-077   & DBQZ4.9 & 10396 & 290.145 & -07.666 & 10.08 & 0.1740 & 200.602 & p & L20\\

1919+145   & DA3.3 & 15280  & 290.418 & +14.678 & 19.80 & 0.0743 & 203.806 & p & L20\\

1935+276   & DA4.2 & 12130  & 294.307 & +27.721 & 17.95 & 0.4361 & 088.686 & p & L20\\

1953-011   & DC6.4 & 7920  & 299.121 & -01.042 & 11.38 & 0.8270 & 212.314 & p & L20\\

2002-110   & DA10.5 & 4800 & 301.395 & -10.948 & 17.33 & 1.0740 & 095.523 & p & L20\\

2007-303   & DA3.5 & 14454  & 302.736 & -30.218 & 17.09 & 0.4280 & 233.492 & p & L20 \\

2008-600   & DC9.9 & 5080  & 303.132 & -59.947 & 16.55 & 1.4276 & 166.100 & p & L20\\

2008-799   & DA8.5 & 5800   & 304.207 & -79.764 & 24.96 & 0.4339 & 127.979 & p & Pi\\

2011+065   & DQ7 & 6400     & 303.481 & +06.712 & 22.37 & 0.6297 & 203.39  & p & Pi \\

2032+248   & DA2.4 & 19983  & 308.591 & +25.063 & 14.65 & 0.6920 & 215.554 & p & L20\\

2039-202   & DA2.5 & 19207   & 310.644 & -20.076 & 21.10 & 0.3672 & 104.831 & p & G11\\

2039-682   & DA3.1 & 15855 & 311.089 & -68.089 & 22.00 & 0.3269 & 144.462 & sp & G11\\

2040-392   & DA4.5 & 10830  & 310.955 & -39.055 & 22.63 & 0.306  & 179     & p &G11 \\

2047+372   & DA3.6 & 14070   & 312.277 & +37.470 & 17.28 & 0.2190 & 047.150 & p& L20 \\

2048+263   & DA9.7 & 5200  & 312.586 & +26.511 & 20.08 & 0.5149 & 235.044 & p & L20\\

2048-250   & DA6.6 & 7630  & 312.749 & -24.867 & 22    & 0.2776 & 129.885 & sp & GBD \\

2054-050   & DC10.9 & 4620  & 314.199 & -04.844 & 17.05 & 0.8020 & 106.562 & p & L20\\

2058+342   & DB4.1  &       & 315.08963 & 34.43917 &-& 0.168 & 42.6 & sp & Ber11\\

2105-820   & DA4.7 & 10620  & 318.320 & -81.820 & 17.06 & 0.5164 & 146.372 & p &L20 \\

2111+072   & DA7.8 &  6470    &  317.370 & +07.451 & 24.1 &  0.341 &   70.289   & sp & Lim\\

2115-560   & DA6 & 9736    & 319.902 & -55.837 & 22.00 & 0.4652 & 115.463 & sp & G11\\

2117+539   & DA3.6 & 13990  & 319.734 & +54.211 & 19.72 & 0.2130 & 336.371 & p& L20  \\

2118-388   & DC9.6 & 5244  & 320.523 & -38.643 & 22.00 & 0.1786 & 112.009 & sp & Sub08\\

2119+040   & DA9.0 &       & 320.55144 & 4.23249 & - & 0.3897 & 28.341 & sp & Say\\

2126+734   & DA3.8 & 16,104   & 321.740 & 73.645  & 21.23 & 0.2915 & 171.119 & p & G11 \\

2133-135   & DA5.0 & 9736  & 324.068 & -13.309 & 20.40 & 0.2935 & 118.486 & sp & Sub08 \\

2138-332   & DZ7 & 7240    & 325.489 & -33.008 & 15.62 & 0.2100 & 228.500 & p & L20\\

2140+207   & DQ6.1 & 8200  & 325.670 & +20.999 & 12.51 & 0.6819 & 199.445 & p & L20\\

2149+021   & DA2.8 & 17353  & 328.105 & +02.388 & 24.50 & 0.3003 & 177.328 & p & G11\\

2151-015   & DA6 & 8400    & 328.526 & -01.285 & 19.60 & 0.2851 & 178.191 & sp & G11\\

2154-512   & DQ8.3 & 6100  & 329.421 & -51.006 & 16.12 & 0.3746 & 184.738 & p & L20\\

2159-754   & DA5.6 & 9040  & 331.086 & -75.223 & 15.62 & 0.2042 & 238.300 & p & L20\\

2210+565   & DA3.0 & 16790  &  332.973 & +56.829 &   18.1 & 0.147   &  217.0     &  sp & Say\\

2211-392   & DA8.1 & 6920  & 333.638 & -38.983 & 18.69 & 1.068  & 109.6   & p & Say\\

2211-392   & DA8.1 & 6920  & 333.644 & -38.985 & 18.79 & 1.0560 & 110.100 & p & Say\\

2215+386   & DC10.6 & 4700 & 334.448 & +37.130 & 25.12 & 0.4691 &  78.69  & p & Pi \\

2226-754   & DC11.9 & 4230 & 337.665 & -75.232 & 12.8  & 1.8680 & 167.500 & sp & L20\\

2226-755   & DC12.1 & 4177 & 337.638 & -75.255 & 14.0  & 1.8680 & 167.500 & sp & L20\\

2246+223   & DA4.7& 10647   & 342.273 & +22.608 & 19.04 & 0.5253 &  83.551 & p & L20\\

2248+293   & DA9 & 5580    & 342.845 & +29.662 & 20.92 & 1.2575 &  83.745 & p & G11 \\

2251-070   & DZ12.6 & 4000 & 343.472 & -06.781 & 08.51 & 2.5718 & 105.600 & p & L20 \\

2253+054   & DA9 & 5600    & 343.982 & +05.755 & 24.46 & 0.4471 & 127.101 & p & Pi\\

2311-068   & DQ6.8 & 7440  & 348.604 & -06.546 & 25.1  & 0.3815 & 244.039 & p & Pi\\

2322+137   & DA10.7 & 4700  & 351.332 & +14.060 & 22.27 & 0.37   &  71.565 & p & L20 \\

2326+049   & DAV4.3 & 12206  & 352.198 & +05.248 & 13.62 & 0.4934 & 236.406 & p & G11\\

2336-079   & DA4.6& 10938   & 354.711 & -07.688 & 15.94 & 0.0500 & 140.000 & p & L20\\

2341+322   & DA4.0 & 13128   & 355.961 & +32.546 & 17.60 & 0.2290 & 252.150 & p & L20\\

2347+292   & DA9 & 5810    & 357.479 & +29.567 & 22.39 & 0.5065 & 185.666 & p & Pi \\

2351-335   & DA5.7 & 8850   & 358.504 & -33.275 & 23.35 & 0.5081 & 219.4   & p & G11\\

2359-434   & DA5.9 & 8648  & 000.544 & -43.165 & 08.16 & 0.8878 & 138.400 & p & L20\\

\hline

\enddata
\tablecomments{\\s - spectrophotometric \\p - trigonometric 
parallax \\a - weighted mean average. \\  \\ 
{\bf {Key to References in Table 1 }}  
\\
G11:	    Gianninas et all (2011)\\ 					   
Lim:	    Limoges et al. (2013) \\ 					   
Say:	    Sayres et al.(2012) \\ 					   
Pi:	    Trig. Parallax 	\\ 				   
DR7:	    Holberg et al. (2013)\\ 					   
Sub08:   Subasavage et al. (2008)\\ 					   
GBR11:   Gianninas Bergeron \& Ruiz (2011)\\ 					   
Sub07:   Subasavage et al. (2007)	\\ 				   
Ber11:   Bergeron et al. (2011)		\\ 			   
GJ: 	    Gliese \& Jahreise (1995)	\\ 				   
Kilic10: Kilic et al. (2010)		\\ 			   
Koe11:   Koester et al. (2011)		\\ 			   
Trem11:  Tremblay et al. (2011)		\\ 			   
} 						   

\end{deluxetable}
						 
\setlength{\hoffset}{00mm} 

\clearpage

\begin{deluxetable}{lcrr}
\tablecaption{Distribution of WD Spectral Subtypes within 25 Parsecs}
\tablewidth{0pc}
\tablenum{2}
\tablehead{
\colhead{Spectral Type} 
          &\colhead{Range of $T_{eff}$}
                             &\colhead{Number of Stars}
                                        &\colhead{\% of total} 
                                                            }

\startdata 
 DA     &       4590 - 25193  &     125  &   54\%\\

DAZ    &         5093 - 20464  &     11 & 5\% \\

DAH/DAP    &  4500 -  30510          &  14     &  6\% \\

Magnetic Non-DA &                     & 5        &       \\

DB     &            16714 &            2  &  0.8\%\\

DBQZ   &            10200 &           1  &  0.4\%\\

DC      &   2600 - 7300  &     28  & 12\% \\

DCP    &              6010     &          3& 1.3\%  \\

DQ     &           5590 - 10900   &      23  & 8\%\\

DQP    &           4948 - 6070    &       3  & 0.9\% \\

DQZ     &               7740      &         1 &  0.4\%\\

DZ      &           4000 - 7500    &     13  & 5\%  \\

DZA    &            7600  - 8440 &     2  & 0.9\%   \\

\enddata

\end{deluxetable}

\begin{deluxetable}{llrrrr}
\tablecaption{Space Motions of White Dwarfs Within 25 pc }\tablewidth{0pc}\tablenum{3}\tablehead{\colhead{WD No.}&\colhead{Type}&\colhead{U}&\colhead{V}&\colhead{W}&\colhead{T} }

\startdata

0000-345 &    DCP9   &   -11.9&    -43.7 &     3.2&     45.4\\

0008+424  &   DA6.8  &   -10.0 &    -1.9 &   -17.4 &    20.2\\

0009+501   &  DAH7.7 &   -28.1 &     8.6 &   -24.1  &   38.0\\

0011-134   &  DCH8.4 &     -75.5 &   -31.1&     -9.8  &  82.3\\

0011-721  &   DA8.0   &   10.3 &   -23.2 &    12.2 &    28.2\\

0029-031  & DA11.3&    68.3  &   -23.0  &     2.6 &      72.1\\

0038+555   & DQ4.6&      39.4  &    -2.0   & -11.9 &     41.2\\

0038-226  &   DQ9.3 &    -24.9 &    -4.6 &    -1.2 &    25.3\\

0046+051   &  DZ8.1  &      -2.8 &   -53.6 &   -30.3 & 61.6\\

0108+048  & DA6.4&       38.2   &    -0.1 &     11.8  &      40.0\\

0108+277   &  DAZ9.6  &  -11.9 &     0.3 &    -9.9  &   15.5\\

0115+159   &  DQ6   &    -21.4 &   -27.2 &   -32.0 &    47.2\\

0121-429   &  DAH7.9 &    -0.7 &   -46.8  &   11.1 &    48.1\\

0123-460   & DA8.5&      24.3  &   -91.0  &    24.9 &     97.4\\

0134+883  &DA2.8&     -11.1   &    5.1  &     6.1 &     13.7\\

0135-052  &   DA6.9 &     16.8&    -37.2 &    -1.6 &    40.9\\

0141-675  &   DA7.8   &    -21.8 &   -22.8 &    21.5 &    38.2\\

   0148+467  &   DA3.8  &     2.6  &    1.8  &    8.3 &     8.8\\

   0148+641  &   DA5.6 &     12.3 &   -11.2  &   -6.8  &   18.0\\

0208+396  &   DAZ7.0  &   43.0 &   -62.9 &    -6.7 &    76.5\\

   0210-510  &   DA10    &   85.6 &   -42.1  &  24.0   &   98.4\\

      0213+427 &    DA9.4  &    37.8 &   -67.3  &  -19.8  &   79.8\\

0227+050   & DA2.7&       -0.1  &    -8.9  &     7.1  &    11.4\\

   0230-144   &  DC9.5  &   -20.7 &   -43.3 &   -13.3  &   49.9\\

0231-054   & DA3.7&      68.5  &    -6.7  &   -56.4 &     89.0\\

  0233-242   &  DC9.3  &   -24.2 &   -34.1 &    -9.8  &   43.0\\

0236+259   &DA9.2&      29.1&    -21.1   &   -8.9 &     37.0\\

0243-026   &DAZ7.4&         8.9   &  -51.6   &  -34.5 &     62.7\\

0245+541   &  DAZ9.7  &   -16.4  &   10.9 &   -24.2  &   31.3\\

0255-705  &DAZ4.7&       22.2  &   -82.0   &    -0.6 &     84.9\\

   0310-688   &  DA3.1   &    0.1  &   -4.2  &    2.8 &     5.1\\

0311-649   &DA4.0&       7.9  &   -12.8  &    9.5 &     17.8\\

0322-019    & DAZ9.7 &   -21.5 &   -66.8 &   -20.7 &    73.2\\

 0326-273   &  DA5.4 &     46.7 &   -29.7&     47.2 &    72.8\\

0341+182   &  DQ7.7   &  -16.0 &   -94.9  &  -23.8 &    99.1\\

0344+014   &  DQ9.9 &     -7.9 &   -43.8  &   -4.9 &    44.8\\

0357+081   &  DC9.2  &   -25.3  &   -4.3 &   -36.8 &    44.9\\

 0413-077   &  DAP3.1  &     -44.2 &   -34.8 &   -73.0 &  92.2\\

0419-487     & DA8&     -1.2  &   -91.1  &   -57.4  &    107.7\\

0416-594  &   DA3.8  &    -4.9&     -5.7&     -1.6  &    7.7\\

0423+044   &DA9&      -6.9  &   -80.5  &    17.2   &   82.6\\

0423+120  &   DA8.2  &       5.3 &    18.1  &    3.8 &   19.3\\

0426+588   &  DC7.1    &     0.9&    -35.4 &    -3.7 &   35.6\\

0431-360   &DA10.0&      10.8   &  -19.6  &   27.0 &    35.1\\

0431-279   &DC9.5 &      11.5  &   -28.8  &    33.3  &    45.5\\

0433+270   &  DA9.3  &    -0.1  &  -19.8  &    7.0  &   21.1\\

0435-088   &  DQ8.0 &    -25.1 &   -63.2 &   -21.6 &    71.4\\

0457-004  &   DA4.7  &    -6.1&    -23.8  &    2.5 &    24.7\\

0503-174   & DAH9.5&      32.2  &   42.5  &    41.0 &     67.2\\

0511+079  &DA7.7&      -14.5 &     -9.2 &   -30.6  &    35.1\\

 0532+414   &DA6.8&       1.2 &    12.4 &    -13.4  &    18.2\\

  0548-001   &  DQP8.3  &     4.6  &    6.3 &    10.5 &    13.1\\

0552-041   &  DZ11.8 &    -25.3 &   -65.7 &   -20.2 &    73.3\\

0553+053   &  DAP8.9 &   -12.2 &   -19.8  &  -31.2  &   38.9\\

0628-020   &  DA     &    -5.2 &   -7.7   &   -17.7  &  20.0 \\

0615-591   &DB3.2&        -11.3   &    -0.1  &     7.1 &  13.3\\

   0618+067   &DA8.1&      -9.0  &   -26.9   &   49.9 &     57.4\\

0620-402 & DZ6&       -13.9   &  -25.0    &  -7.2 &     29.5\\

0627+299  &DA5&        -0.80   &   -1.4  &     -0.6 &      1.7\\

0628-020  &DA&  61.0  &   -34.5  &   -27.2  &  
75.2\\

0642-166   &  DA2  &      -0.6 &   -10.2 &   -14.5  &   17.7\\

 0644+025   &  DA8    &     9.4 &    15.1 &   -30.9 &    35.7\\

0644+375   &  DA2.5  &   -19.7  &  -39.6 &   -32.7  &   55.1\\

0651-398A  &DA7.0&        5.2  &    28.7 &     8.9&     30.5\\

   0655-390  &   DA8.0     &      2.7 &     2.3 &  -28.2 &   28.4\\

0657+320    & DC10.1  &    -31.7   &  -52.9  &    13.2 &     63.1\\

0659-063   & DA7.8  &  -19.7 &  -37.5  & -29.5  &  51.6\\

0708-670  & DC &          4.7 &    4.9&   -19.9 &   21.1\\

0706+377  &DQ7.6&       -7.2  &   -13.0 &    -31.6 &     34.9\\

0727+482.1 & DA10.0 &  -14.6 &  -40.5 &  -14.7  &  45.5\\

0727+482.2 & DA10.0 &  -14.9  & -41.1  & -14.9   & 46.3\\

0728+642  & DAP11.2 &      -4.5  &  -9.2 &    3.9  &  11.0\\

0736+053  &  DQZ6.5  &   0.6  & -12.0 &  -18.4 &   21.9\\

0738-172  &  DAZ6.7  & -30.5 &  -29.2  &  30.2 &   51.9\\

0743-336   & DC10.6   &    49.4 &   67.6  &  61.4 &  103.9\\

0747+073.1 & DA12.1    &  -76.7&  -126.3  & -49.0 &  155.7\\

0747+073.2  &DA11.9  &    -76.7  &-126.3  & -49.0  & 155.7\\

0749+426   &  DC11.7 &   -17.8 &   -29.6  &    5.6  &   35.0\\

0751-252   & DA10.0 &  19.1  &     15.5 &   -13.8 &    28.3\\

0752-676   &  DC10.3 &   -29.3 &   -15.4   &  13.6  &   35.9 \\

0753+417   & DA7.3   &  -13.8  &    -28.6  &   -3.5 &    31.9 \\

0806-661   &  DQ4.2   &    -17.2  &   -6.5  &    7.3  &   19.8 \\

0805+356   & DA7.3   &    0.5    & -4.9  &    -6.6   &    8.3 \\

0810+489   & DC6.9  &     -9.0 &   -15.1  &     5.0  &    18.3\\

0810+489   &DC6.9&     -14.5  &   -22.8   &    8.6   & 28.3\\

0816-310   &DZ7.6&     -45.0    & -52.7  &   -37.0  &  78.6\\

0821-669   &  DA9.8  &    16.8   &   4.5  &    4.4   &  18.0\\

0827+328  &DA6.9&        4.1  &   -45.4  &    -6.6 &     46.1\\

0839-327  &   DA5.3   &   38.2 &    26.6 &     4.6 &    46.8\\

0840-136  &   DZ10.3 &    14.0  &   -0.0  &  -21.0  &   25.2\\

0843+358  &DZ6&          7.6  &    -5.5 &    -13.6&     16.5\\

0856+331  &DQ5.1&       21.4  &      0.883  &   -24.3 &     32.4\\

0912+536   &  DCP7    &    21.0  &  -42.5  &  -24.3  &   53.3\\

0946+534   &DQ6.2&      20.2 &     -7.3   &  -16.1 &     26.9\\

0955+247  &   DA5.8  &     6.1  &  -28.1 &   -17.7  &   33.8\\

0955+247   &DA5.8 &      19.1  &   -40.3  &   -11.8   &   46.2\\

0959+149  &DC7&      -39.4   &   56.1  &  -127.2  &     144.5\\

1009-184  &   DZ7.8     &    36.1 &    -8.8 &   -26.4  & 45.6\\

1009-184   & DZ8.5 &      35.1  &    -6.0   &  -26.8 &     44.5\\

1012+083.1  &DA7.8&     38.4   &   10.1  &   -12.2  &    41.5\\

1019+637&     DA7.3  &   -14.7 &    17.9 &     3.5   &  23.4\\

1033+714  &   DC9  &     138.8 &   -73.2 &   -58.0  &  167.3\\

1036-204  &   DQP10.2 &   36.0 &    17.9 &    20.7  &   45.2\\

1043-188  &   DQ8.1  &   110.7  &  -71.1 &  -107.8 &   170.1\\

1055-072  &   DA6.8 &     42.3 &   -10.4 &   -16.3 &    46.5\\

1105-048  &DA3.5&      -17.3  &   -37.8  &   -29.8 &     51.2\\

1116-470  &   DC &        24.3  &   -9.2 &   -7.0 &    27.0\\

1121+216   &  DA6.7  &    57.8&    -25.3 &   -21.5 &    66.7\\

1124+595   &  DA4.8  &   -12.2  &    1.3   &   6.0 &    13.7\\

1132-325  &   DC   &     -11.4 &    22.5  &   35.8 &    43.8\\

1134+300   &  DA2.5   &    9.2  &   -4.6  &   -2.9  &   10.7\\

1142-645   &  DQ6.4   &  -52.1 &    25.3&      6.6  &   58.3\\

1149-272   & DQ8.1&      24.9  &    -8.5     &  -0.8 &     26.4\\

1202-232   &  DAZ5.8   &    3.0  &    6.1  &    8.6 &    11.0\\

1208+576  &DAZ8.6&        -45.5 &    -10.6  &    23.0 &     52.0\\

1214+032   &DA8.0&      64.0  &   -11.2  &     2.3 &    64.9\\

1223-659   &  DA6.5    &      0.7 &     0.1  &   -8.4  &    8.4\\

1236-495   &  DA4.4  &    24.0  &  -17.0   &  -8.6  &   30.7\\

1241-798   & DC/DQ&      38.1   &  -36.9    &  33.3  &    62.6\\

1242-105   &DA6.3&      13.8  &   -44.3  &    24.3 &     52.4\\

1257+037   &  DA8.7  &    -6.2 &   -69.4  &  -25.6 &    74.2\\

1309+853   &  DAP9   &   -16.3  &   -1.3   &  15.4   &  22.5\\

1310+583   &DA4.8&     -16.3  &     6.7  &    5.7 &     18.5\\

1310-472   &  DC11.9  &   -133.6 &    81.4 &   -51.5 &   164.8\\

1315-781   &  DA    &    -23.5  &   +26.4  &  -32.3  &  47.8 \\

1315-781  &DC8.8&      -30.2  &    34.0   &  -41.7 &     61.7\\

1327-083   &  DA3.7  &    49.2 &   -76.6  &   -7.4  &   91.3\\

1334+039   &  DZ10.0 &    87.5 &  -122.2  &    8.5  &  150.5\\

1337+705  &DAZ2.8&       43.7  &   -15.5  &   29.9 &     55.1\\

1339-340  &DA9.5&         55.0  &   108.0  &   177.4 &    214.8\\

1344+572  &DA3.8&       27.3  &    26.8  &    37.5  &     53.5\\

1344+106   &  DAZ7.1   &     54.6  &  -62.3  &   14.1 &    84.1\\

1345+238   &  DA10.7  &   67.0 &   -47.4  &   20.1  &   84.5\\

1350-090   &DAP5&     -21.7   &  -23.0    & -23.2  &    39.2\\

1401+457   &DC19&      16.9 &    -27.6   &    5.5 &     32.9\\

1403+451   &DB&       3.9    &  -2.1 &      1.2 &      4.5\\

1425-811  & DAV4.2&         -7.9  &   -26.2  &   -44.0 &      51.779\\

1436-781  &DA8.1&       29.7   &  -32.5   &   23.3 &     49.8\\

1444-174   &  DC10.1  &   33.0 &   -61.4 &    17.6 &    71.9\\

1529+141   & DA9.6&     -10.6  &   -13.1    &  -7.2 &   18.3\\

1532+129    &DA9.6&     -1.8  &   -26.2   &    3.9 &     26.5\\

1538+333   &DA5.6&      12.0    &  -5.2   &   11.8  &    17.6\\

1544-377   &  DA4.7  &     5.5 &   -22.4  &    7.6 &    24.3\\

1609+135   &  DA5.8  &   -25.2 &   -35.3 &   -17.4 &    46.8\\

1620-391   &  DA2     &   -1.3  &    2.8  &   -3.2  &    4.5\\

1625+093   &DA7.3 &     -23.9   &  -46.7  &   -12.3 &     53.9\\

1626+368  &   DZ5.5  &    35.1  &   15.5 &    35.6 &    52.3\\

1632+177   &  DA5   &      -2.7 &     2.3 &    -5.3   &   6.4\\

1633+433  &   DAZ7.7  &  -12.7 &    -4.2  &  -13.9  &   19.4\\

1633+572   &  DQ8.2  &      45.4&     -3.0&     54.5 &    71.0\\

1639+537   & DAH6.7&      -3.6   &  -19.1    &   8.9 &     21.4\\

1647+591   &  DAV4.2  &   -6.0  &   -3.0  &   -5.4  &    8.6\\

1655+215  &   DA5.4  &   -25.9  &  -35.9  &  -18.4 &    47.9\\

1655+215   &DA5.4&     -38.0  &   -39.0  &   -17.7 &      57.3\\

1658+440   &DAP1.7&      -6.6   &  33.0  &    28.2 &     43.9\\

1705+030  &   DZ7.1 &    -16.6 &   -23.1 &   -13.4 &    31.5\\

1729+371   &DA7.3&     -28.2  &    21.2  &    13.8 &     37.9\\

1748+708   &  DQP9.0 &      5.8 &   -10.8 &    34.2  &   36.3\\

1756+143   &DA9.0&     -36.3   &  -86.3 &     49.1 &    105.7\\

1756+827  &   DA7.1  &     8.8 &   -30.3 &   106.4 &   111.0\\

1814+134  &   DA9.5  &   -47.7   & -72.0 &    -8.6  &   86.8\\

1817-598    &DA5.8&    -15.1  &   -22.9  &     2.3  &    27.6\\

1820+609  &   DA10.5 &    -9.8 &    -9.0 &   -19.7 &    23.7\\

1829+547  &   DQP7.5  &    2.8 &    -0.7 &    24.3 &    24.4\\

1840+042   &&      32.0   &  -23.6 &     19.9  &   44.4\\

1900+705   &  DAP4.5 &     6.7 &     5.5  &    6.0 &    10.5\\

1917+386    & DC7.9  &    -5.2  &   -5.7 &    -8.6  &   11.6\\

1917-077   &  DBQA5  &    -5.4 &    -6.4 &    -0.5 &     8.4\\

1919+145   &  DA3.5  &    -4.2 &    -5.0 &    -0.7 &     6.6\\

1935+276   &  DA4.5  &    16.0  &   10.2&    -31.7 &    37.0\\

1953-011  &   DAP6.5  &     -32.3 &   -31.9  &    3.7 &    45.6\\

2002-110 &    DA10.5  &   40.4   &   9.8  &  -76.6  &   87.2\\

2007-303   &  DA3.3 &    -22.0  &  -18.3  &   18.0  &   33.8\\

2008-600   &  DC9.9 &    -17.7  &  -62.3  &   -3.3  &   64.9\\

2008-799    & DA8.5&     18.0    &  -6.5   &  -20.7    &  28.2\\

2011+065  &DQ7&      -64.2   &  -24.8  &   -17.0  &    70.9\\

2032+248   &  DA2.5  &   -36.9 &   -24.9 &    -2.8  &   44.7\\

2039-202  &DA2.5&       15.9  &    -3.9 &    -30.9 &     35.0\\

2040-392  &DA4.5&      -13.6  &   -25.2  &     -0.3  &  28.7\\

2039-682   &DA3.1&        0.698  &   -17.8  &    -9.9 &     20.4\\

2047+372   &  DA4  &      13.4 &     5.5 &    -1.4 &    14.6\\

2048+263   &  DA9.7  &   -37.6 &   -16.2  &   11.8 &    42.7\\

2048+263    &DA9.7&    -79.5    & 104.1    & -13.4 &    131.7\\

2048-250   &DA6.6&       5.3  &   -13.6  &  -20.0  &    24.8\\

2054-050  &   DC10.9  &   31.1  &  -11.4 &   -54.1 &    63.5\\

2105-820  &   DAP4.9  &    12.5 &   -17.4 &     3.6  &   21.8\\

2117+539   &  DA3.5  &    -0.8  &    2.1 &    18.3 &    18.5\\

2115-560   &DA6&      17.5  &   -18.6  &   -31.4   &   40.5\\

2118-388    &DC9.6&      9.3   &   -6.3   &  -11.6 &     16.1\\

2126+734   &DA3.8&       3.2 &    11.8  &   -23.5  &   26.5\\

2133-135   &DA5.0&      11.7    & -12.8  &   -20.3 &   26.7\\

2138-332   &  DZ7  &     -15.0 &    -8.3  &    8.4  &   19.1\\

2140+207   &  DQ6.1  &     -28.0 &   -18.9 &    -17.1 &    37.9\\

2149+021   & DA2.8&       -28.4  &    -3.2 &   -37.7  &    47.3\\

2151-015   & DA6&     -59.0   &   52.4  &   -85.8 &     116.6\\

2154-512   &  DQ7   &    -12.4 &   -22.4  &    9.9  &   27.5\\

2159-754   &  DA5   &    -27.6 &     7.6  &   18.5 &    34.1\\

2211-392   &  DA8   &     57.8  &  -43.7  &  -44.5  &   85.1\\

2207+142   &DA6.6&      21.1   &   83.8 &    -45.7  &    97.7\\

2209-147   &DA6.6&      54.9  &   -21.5  &    30.4   &   66.3\\

2211-392   &DA8.1&      46.6  &   -43.1  &   -61.1 &     88.2\\

2215+386   &DC10.6&      48.7  &    -7.1   &  -20.6 &     53.3\\

2226-754   &  DC9.9  &    -0.5 &   -88.5  &   71.5  &  113.8\\

2226-755   &  DC12.1  &   -0.5 &   -88.5  &   71.5  &  113.8\\

2246+223   &  DA4.7  &    42.4  &  -10.4  &  -16.9  &   46.8\\

2248+293  &DA9&      111.2  &   -29.8  &   -43.0 &   122.9\\

2251-070   &  DZ13 &      68.0 &   -48.2 &   -50.0 &    97.2\\

2253+054   &DA9&      20.3 &    -33.8  &   -34.2  &      52.2\\

2311-068   &DQ6.8&       -45.3  &    -1.7  &     7.4 &     45.9\\

2322+137    & DA10.7   &   3.2   &  -0.5  &   -0.3  &    3.2\\

2326+049   &  DAZ4.4  &  -31.6   &  -2.1  &   -1.3   &  31.7\\

2336-079    & DAZ4.6   &  -5.8  &  -13.4 &    -5.9 &    15.8\\

2341+322   &  DA4.0 &    -18.8  &    5.6   &  -0.4   &  19.6\\

2347+292   &DA9&     -18.3  &    2.7   &  -57.2 &     60.1\\

2351-335    &DA5.7&    -62.0    & -22.5  &   -38.2  &    76.2\\

2359-434   &  DAP5.8  &    13.9  &  -16.4 &   -27.6  &   35.0\\

\enddata
\end{deluxetable}

\begin{deluxetable}{llrrrr}
\tablecaption{
Kinematical Statistics of Spectroscopic Subgroups Within 25pc}
\tablewidth{0pc}
\tablenum{4}
\tablehead{
\colhead{Type}
     &\colhead{Number}
             &\colhead{$<U>$}
                   &\colhead{$<V>$}
                           &\colhead{$<W>$}
                                  &\colhead{$<T>$} } 
\startdata   

DA/DAZ     &   133 &   2.5 & -18.4 &  -5.8 & 48.9 \\
 
           &       &  33.9 &  33.8 &  28.6 & 33.0 \\ 

DQ         &    15 &  -2.4 & -16.0 &  -3.3 & 46.8 \\ 
           &       &  38.7 &  26.8 &  25.2 & 22.2 \\

DZ         &    14 &   2.0 & -21.5 & -14.0 & 47.0 \\ 
           &       &  35.0 &  20.4 &  25.3 & 25.0 \\ 

DC         &    26 &  -4.1 & -27.0 &  -7.8 & 66.6 \\ 
           &       &  46.7 &  49.8 &  41.3 & 50.8 \\ 

DAP/DAH    &    11 &   3.3 & -10.2 &   0.9 & 42.0 \\ 
           &       &  30.6 &  30.7 &  26.3 & 27.5 \\ 

Mag.non-DA &     5 &   4.2 & -19.5 &  -0.4 & 47.0 \\ 
           &       &  25.4 &  28.5 &  16.9 & 25.0 \\ 

\enddata

\end{deluxetable}

\end{document}